\documentclass{ecai} 

\usepackage{xspace}
\usepackage{latexsym}
\usepackage{amssymb}
\usepackage{amsmath}
\usepackage{amsthm}
\usepackage{booktabs}
\usepackage{enumitem}
\usepackage{graphicx}
\usepackage{color}
\usepackage{cite}
\usepackage{multirow}
\usepackage{cite}
\usepackage{amsmath,amssymb,amsfonts}
\usepackage{algorithmic}
\usepackage{graphicx}
\usepackage{textcomp}
\usepackage{xcolor}
\usepackage{comment}
\usepackage{subcaption}
\usepackage{booktabs}
\usepackage{threeparttable}
\newcommand*{\revise}[1]{\textcolor{black}{#1}}
\newcommand*{\revises}[1]{\textcolor{black}{#1}}
\newcommand*{\note}[1]{\textcolor{black}{#1}}

\newcommand{\sysname}{{DHLight}\xspace}

\newcommand{\BibTeX}{B\kern-.05em{\sc i\kern-.025em b}\kern-.08em\TeX}

\begin{document}


\begin{frontmatter}




\title{\sysname: Multi-agent Policy-based Directed Hypergraph Learning for Traffic Signal Control}


\author[A]{\fnms{Zhen}~\snm{Lei}}
\author[A]{\fnms{Zhishu}~\snm{Shen}\footnote{Corresponding Authors. Email: z\_shen@ieee.org (Zhishu Shen) and tiehuaz@tongji.edu.cn (Tiehua Zhang).}}
\author[A]{\fnms{Kang}~\snm{Wang}} 
\author[B]{\fnms{Zhenwei}~\snm{Wang}} 
\author[C]{\fnms{Tiehua}~\snm{Zhang}\footnotemark} 

\address[A]{School of Computer Science and Artificial Intelligence, Wuhan University of Technology, China}
\address[B]{School of Computer Science, University of Nottingham Ningbo China, China}
\address[C]{School of Computer Science and Technology, Tongji University, China}


\begin{abstract}
Recent advancements in Deep Reinforcement Learning (DRL) and Graph Neural Networks (GNNs) have demonstrated notable promise in the realm of intelligent traffic signal control, facilitating the coordination across multiple intersections. However, the traditional methods rely on standard graph structures often fail to capture the intricate higher-order spatio-temporal correlations inherent in real-world traffic dynamics. Standard graphs cannot fully represent the spatial relationships within road networks, which limits the effectiveness of graph-based approaches. In contrast, directed hypergraphs provide more accurate representation of spatial information to model complex directed relationships among multiple nodes. In this paper,  we propose \sysname, a novel multi-agent policy-based framework that synergistically integrates directed hypergraph learning module.  This framework introduces a novel dynamic directed hypergraph construction mechanism, which captures complex and evolving spatio-temporal relationships among intersections in road networks. By leveraging the directed hypergraph relational structure, \sysname empowers agents to achieve adaptive decision-making in traffic signal control. The effectiveness of \sysname is validated against state-of-the-art baselines through extensive experiments in various network datasets. \revises{ We release the code to support the reproducibility of this work at 
https://github.com/LuckyVoasem/Traffic-Light-control}%
\end{abstract}

\end{frontmatter}


\section{Introduction}
\label{sec:intro}
Traffic Signal Control Systems (TSCSs) are essential components of contemporary transportation infrastructure, playing a vital role in managing and coordinating the flow of traffic across road networks. Traditional traffic signal control methods rely on presetting light timings based on historical traffic patterns, periodically adjusted to manage vehicle flow through intersections~\citep{de2020development}. However, the substantial rise in vehicle numbers has introduced challenges such as a lack of adaptability to dynamic traffic conditions and rudimentary delay coordination among traffic lights within the network. As a result, these systems fail to effectively mitigate traffic congestion throughout the road networks~\citep{0Variational,NOAEEN2022116830}.



In recent years, the integration of Artificial Intelligence (AI) and Internet of Things (IoT) technologies has been explored to enhance intelligent traffic light control systems. Leveraging IoT sensor data from nodes across the road networks, coupled with advanced big data analytics, these intelligent systems can autonomously adjust the timing of traffic signal phases to reflect real-time traffic conditions. This approach significantly improves the responsiveness and efficiency of traffic management~\citep{mao2022comparison}, which enhances road network throughput and alleviates the pressure of traffic congestion. Additionally, a line of research \citep{yau2017survey,intell2018,2019Time} focuses on using Deep Reinforcement Learning (DRL) to tackle the complex situation of road networks, aiming to model the traffic environment and achieve intelligent traffic light control. 
By using reward feedback, DRL progressively learns to make optimal decisions for signal light phase selections within the road networks. Some classic DRL algorithms such as Deep Q-Network (DQN) and Asynchronous Advantage Actor Critic (A3C) have been implemented in the traffic control use case with promising results~\citep{bouktif2023deep,su2023emvlight}. However, these methods primarily focus on optimizing individual intersections (single-agent) without considering the interactions between intersections, which may result in local optimal solutions~\citep{2020STMARL}. In contrast, multi-agent reinforcement learning enables multiple traffic lights to collaborate and learn cooperative strategies. For instance, the work in \citep{wang2021adaptive} merged the cooperative vehicle infrastructure system into a value-based multi-agent reinforcement learning model and designed a spatial discount reward, facilitating collaborative strategy optimization among traffic lights. Nonetheless, due to the complexities of the large action space dimensions, achieving the maximum reward value is still challenging.

On the other hand, recent advancements in Graph Neural Networks (GNNs) have improved the representation embeddings of traffic signal control nodes by utilizing the inherent topological structure of the road network~\citep{rahmani2023graph}. Inspired by the idea of incorporating the reward feedback mechanism into the learning process, some recent studies have employed spatio-temporal graph neural network methods combined with multi-agent reinforcement learning to reduce vehicle travel time and alleviate congestion in the road network, yielding promising results~\citep{GCN2018,Colight,zhong2021probabilistic}. However, these approaches do not comprehensively account for the impact of spatio-temporal traffic flow between adjacent intersections. Meanwhile, compressing complex relationships into pairwise relationships loses high-order spatial-temporal information, posing challenges in handling real-world dynamic road network situations. To address this issue, hypergraph learning offers a promising approach to capture the complex spatio-temporal relationships among neighboring intersections~\citep{Hypergraph}. In contrast to the standard graph in pairwise style, hypergraphs utilize hyperedges to connect more than two nodes to form hyperedges, by which provides distinct benefits in modeling complex relationships among multiple nodes and enable comprehensive higher-order network analysis~\citep{DirectedHA,cao2024spatial}. However, hypergraphs may overlook the directed nature of spatio-temporal relationships, which limits their effectiveness in accurately capturing dynamic interactions in real-world scenarios like TSCSs. 


To this end, we propose a novel multi-agent policy-based directed
hypergraph learning \sysname for mining high-order hidden correlations among traffic lights in the road network. Compared to the traditional undirected hypergraphs, directed hypergraphs are superior in modeling the directionality of traffic flow and have a distinct advantage in representing complex directed relationships among multiple nodes exist in TSCSs. Moreover, we utilize the Multi-agent Proximal Policy Optimization (MA-PPOs~\citep{PPO}), which employs a simple yet effective clipping mechanism that restricts the magnitude of policy updates, preventing significant policy changes during the learning process. In large-scale complex environments, MA-PPO is an effective solution among DRL algorithms~\citep{ppoIot2023}. The main contributions of this paper are as below: \begin{itemize}
    \item  \textbf{Framework:} We propose a multi-agent policy-based directed hypergraph learning framework. In this framework, agents are deployed at each intersection to coordinate traffic flow conditions between multiple intersections. Interaction between agents is realized by dynamically constructing directed hypergraphs. Empowered by with MA-PPO, signal light decision optimization between multiple intersections strikes a balance between network-level and single-intersection-level. 
    
    \item  \textbf{Directed hypergraph learning}: The directed hypergraph module is introduced to extract the directed spatial attributes of the road network according to the driving direction of traffic flow. 
     Spatial-temporal correlations among multiple node groups are captured through hyperedges generated by a couple of dynamical headset and tailset, which contribute to updating node embeddings via the multi-head attention mechanism.
     
     \item \textbf{Experiments}: We conduct rigorous experiments based on two real-world road network datasets and \note{two synthetic network datasets.} The obtained results validate the effectiveness of our proposal against several state-of-the-art methods in terms of average travel time.
\end{itemize}

The rest of this paper is organized as follows: Section~\ref{sec:related} summarizes the related work. Section~\ref{sec:form} formulates the traffic signal control as a multi-agent reinforcement learning problem, following by the description of our proposed multi-agent policy-based directed hypergraph learning framework in Section~\ref{sec:framework}. Section~\ref{sec:experiment} evaluates the performance of our proposed \sysname. Finally, the paper is concluded with future work in Section~\ref{sec:conclusion}.
\section{Related Work}
\label{sec:related}

\subsection{Reinforcement Learning for TSCSs}
Traditional TSC methods use deterministic offline strategies such as phase switching with predefined rules~\citep{Koonce2008TrafficST,haydari2020deep,Varaiya2013}. In contrast, reinforcement learning methods seek to adjust the signal light switching strategy adaptively based on the real-time situation at the intersection~\citep{NOAEEN2022116830}. \revises{Although these DRL methods can develop effective optimization strategies for individual intersections, the in-depth collaborative interaction between intersections remains inadequately explored.} Therefore, when considering the overall road network, it may become trapped in local optimal  situations. To mitigate this issue, max-pressure control \citep{boukerche2021novel,WeiHuaKDD19,ChenAAAI2020}, independent Advantage Actor-Critic (A2C) \citep{chu2019multi},  policy gradient method \citep{2020Multi}, MixLight \citep{MixLight}, and GPLight~\citep{GPLight} are employed with multi-agent DRL schema, leading to a collaboration of various intersections for the global traffic optimization. However, the aforementioned studies primarily focus on extracting information from the first-order neighbors of each intersection, which overlooks the potential essential high-order information across different intersections.

\subsection{Integration of Reinforcement Learning and Graph Learning}
Due to the inherent topological properties of road networks, our proposed framework is closely related to the recent advances of GNN-based studies~\citep{Colight,GCN2018,yoon2021transferable,zhang2023learning,emarlin+,wang2025TMC}. Therein, Colight~\citep{Colight} utilizes Graph Attention Networks (GAT) combined with parameter-sharing to balance the influences among adjacent intersections. Nishi \textit{et al.} \citep{GCN2018} demonstrated a model that uses Graph Convolutional Networks (GCN) to learn representations among multiple adjacent nodes. GPLight~\citep{GPLight} aggregates neighborhood features from distant intersections using GCN and a grouping model. { eMARLIN+~\citep{emarlin+} achieves the sharing of embedded information among adjacent intersections through an encoder-decoder structure. Wang \textit{et al.}~\citep{wang2025TMC} proposed a spatio-temporal hypergraph-based multi-agent reinforcement learning framework, which integrates hypergraph learning~\citep{TAI} into the multi-agent soft actor-critic algorithm to optimize traffic signal control.}  Although these methods utilize the message-passing strategy in graph learning to absorb the representation information of adjacent nodes, the traffic flow direction on the graph are not incorporated. Comparatively, we present a directed hypergraph-based framework to enhance high-order correlation extraction among multiple nodes in the road networks.

\section{Problem Formulation}~\label{sec:form}
\begin{figure*}[ht]
\centering
\includegraphics[width=7.2in]{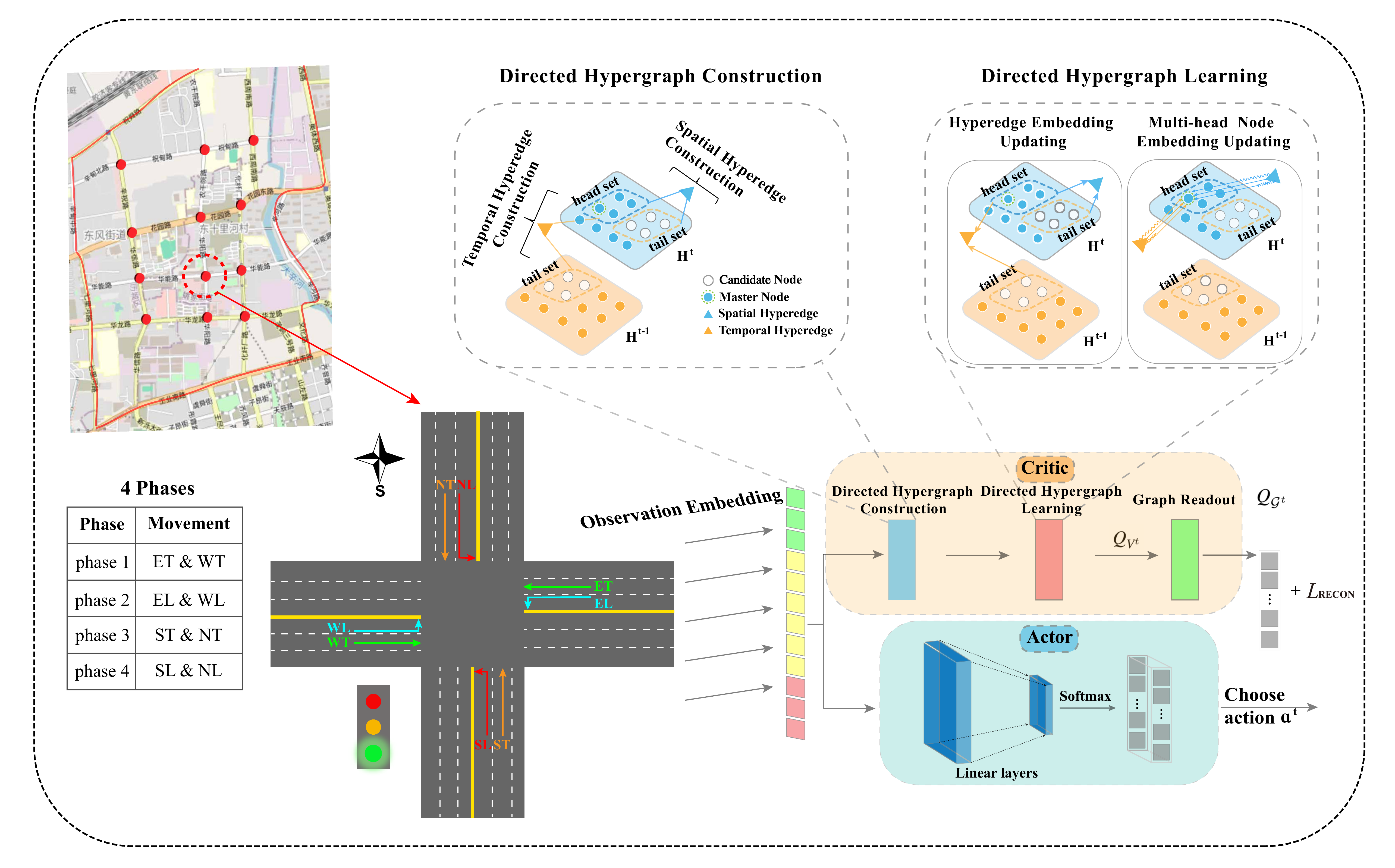}
\caption{An overview of our proposed framework \sysname.}
\vspace {0.2cm}
\label{fig:framework}
\end{figure*}

In this paper, we assume a road traffic scenario featuring three vehicle lanes and four directions of movement: East (E), South (S), West (W), and North (N). A traffic light is deployed at each intersection to regulate traffic flow by managing three types of movements: Drive Through (T), Turn Left (L), and Turn Right (R), achieved through a sequence of different phases. A phase consists of a set of vehicular movement signals that do not conflict with one another. \figurename~\ref{fig:framework} illustrates a road intersection with four distinct phases: Phase 1 (ET and WT), Phase 2 (EL and WL), Phase 3 (ST and NT), and Phase 4 (NL and SL). Vehicles making a right turn at this intersection can proceed during any of these four phases. An agent is employed to manage the phase changes at each intersection. At any given time $t$, the agent assesses the current intersection conditions and chooses one of the four available phases.

This paper frames the control of traffic signal lights as a multi-agent reinforcement learning problem, which deploys an intelligent agent at each intersection to manage the switching of signal light phases. To model the relationship between the traffic state and control action, it is reasonable to simplify the traffic signal control as the Markov decision process, and the
effectiveness is supported by numerous empirical studies\citep{intell2018}.
The problem is characterized by the following components:

\textbf{State space $\mathcal{S}$ and Observation space $\mathcal{O}$}: At time $t$, each agent $i$ can only access a local observation $o_{i}^t$ from local system state $s_{i}^t$.  $o_{i}^t$ consists of agent information, its current phase represented by a one-hot vector, the number of vehicles on each lane connected with the intersection. 
     
\textbf{Action} $\mathcal{A}$: We assume each intersection contains assumed four signal phases. 
At each time $t$, each agent $i$ chooses one of the four traffic signal phases as its action $a_{i}^t$. The decisions made by the traffic signal at an intersection can impact the traffic flow in the surrounding area, which subsequently affects the observations of the agent.

\textbf{Reward $\mathcal{R}$ and Policy $\pi$} : After agent $i$ select the action, it obtains a reward $r_i^t$ by reward function $\mathcal{S}\times\mathcal{A}_1\times \cdots \times\mathcal{A}_N\rightarrow \mathcal{R}$. We define the immediate reward for agent $i$ as: \begin{equation}
    r_i^t = \text{-} \sum_l{u_{i,l}^t}
\end{equation} where $u_{i,l}^t$ is the queue length on the approaching lane $l$  at time $t$.At each intersection, the waiting time for vehicles is affected by the queue length in the lane. As the queue becomes longer, the waiting time for vehicles increases, leading to an overall increase in travel time. Agent $i$ chooses an action $a$ based on a certain policy $\mathcal{O} \times \mathcal{A} \rightarrow \pi$, 
with the goal of reducing travel time for all vehicles within the road network. 

\section{Multi-agent Policy-based Directed Hypergraph Learning Framework}
\label{sec:framework}
In this section, we introduce our proposed framework \sysname for TSCSs. This section first gives an overview of \sysname. Then, a detailed description of the two key modules is presented: directed hypergraph module (including hyperedge construction and directed hyperedge learning) and multi-agent deep reinforcement learning module. The time complexity analysis is provided at the end of this section.

\subsection{Overview}
\label{ssec:framework}
We propose a novel framework \sysname to extract the directed spatio-temporal attributes of the road networks. As illustrated in \figurename~\ref{fig:framework}, the overview of our proposed \sysname is summarized as below:

Firstly, we obtain real-time intersection status from the agents deployed at each intersection as the input to the network\revises{, which refers to the number of vehicles in each lane of the intersection and the encoding of the current phase of the traffic lights.} Subsequently, we employ an on-policy gradient reinforcement learning method PPO, to continuously optimize decision-making at each intersection, thereby enhancing policy performance, as depicted in the bottom right corner of \figurename~\ref{fig:framework}. The PPO composes of an actor network and a critic network. In the actor network, by using the intersection's status, we calculate the probability distribution of each action, then stochastically select an action based on this distribution for execution, thereby enabling the phase transition at the current intersection. Meanwhile, in the critic network, to effectively capture spatio-temporal relationships across different intersections, we utilize the directed hypergraphs to enhance coordination among multiple agents at these intersections. As illustrated in the top right section of \figurename~\ref{fig:framework}, we introduce a directed hypergraph module (comprising  \textit{directed hyperedge construction} and \textit{directed hyperedge learning}), aimed at capturing the complex \note{and directed} spatio-temporal relationships among multiple nodes in traffic signal control, thus updating the representation of each node. Finally, the critic network optimizes the actor network process to enhance decision-making efficiency, thereby improving the overall policy performance of TSCSs. \revises{Moreover, when dealing with large-scale road networks, a partitioning approach can be adopted to divide them into multiple sub-networks, with each sub-network being controlled by \sysname respectively.}

\subsection{Directed Hypergraph Construction}
\label{sssec:DirectedHypergraph}

We assume that the spatial-temporal hypergraph at time $t$ is represented as $\mathcal{G}^{t} = \left\{\mathcal{V}^{t},\mathcal{E}^{t}\right\}$. To account for the temporal relationships among nodes, we incorporate the features of nodes from the previous timestamp to capture historical data following the method proposed in \citep{zhang2023adaptive}. Consequently, the node set $\mathcal{V}^{t}$ includes nodes from the two consecutive timestamps, defined as $\mathcal{V}^{t} = \left\{\emph{V}^{t},\emph{V}^{t-1}\right\}$. The feature matrices for these timestamps are denoted as $\emph{X}^{t}$ and $\emph{X}^{t-1}$, respectively.

A hyperedge encompasses a specific group of nodes that share common attributes or implicit data relationships, representing local group dynamics among the nodes. To facilitate the creation of hyperedges, we introduce two types of nodes: \textit{master nodes} and \textit{candidate nodes}. A master node $\dot{v}\in\mathcal{V}$ serves as a reference point when forming the hyperedge $e\left(\dot{v}\right)\in\mathcal{E}$, which is formed in conjunction with a collection of candidate nodes $S\left(\dot{v}\right) = \left\{\hat{v}\right\}$. We propose a dynamic learning framework that generates two categories of hyperedges: the \textit{spatial hyperedges}, which capture heterogeneity by encoding relationships among different channels at a single timestamp, and the \textit{temporal hyperedges}, which examine interactivity by modeling the ongoing interactions of channels across consecutive timestamps.

Due to the complex spatio-temporal nature of traffic flow, intersections exhibit intricate spatio-temporal correlations. Unlike edges in standard graphs that only describe relationships between two nodes, directed hyperedges in directed hypergraphs can capture the complex spatio-temporal correlations between two sets of nodes, i.e., head node set and tail node set~\citep{DirectedHA}.

Inspired by the methodology proposed in \citep{HGNN+} designed for hypergraph construction, we
adopt the dynamic construction of directed hypergraphs to capture the spatio-temporal correlations between multiple sets of intersections by:
\begin{equation}
    h_i^t = \sigma(o_i^tW_e + b_e)
\end{equation} 
where $o_i^t \in \mathbb{R}^{d}$ is the observation of agent $i$ at time $t$, \revises{ which incorporates the traffic flow state of the intersection and the traffic signal phase.} $W_e \in \mathbb{R}^{d \times l}$ and $b_e \in \mathbb{R}^{l}$ are trainable weight matrix and bias vector. $\sigma(\cdot)$ is the ReLU function. $h_i^t \in \mathbb{R}^{l}$ represents the traffic information of the $i$-th intersection at time $t$. ${H}_{\mathcal{G}}^{t}=\{h_1^t ,..., h_N^t\} \in R^{N \times l} $ is the initial node embedding of the entire road network.

Then, we capture the complex spatio-temporal correlations between nodes in the road network by dynamically constructing spatial hyperedges and temporal hyperedges. The spatial hyperedge $e_{spa}\left(\dot{v}_{i}^{t}\right)\in\mathcal{E}^{t}$ can be generated based on the reconstruction of the master node $\dot{v}^{t}_{i}\in\emph{V}^{t}$ and the spatial candidate  node set $\Tilde{S}^{spa}\left(\dot{v}^{t}_{i}\right) = \left\{v|v\in\emph{V}^{t},v\neq\dot{v}^{t}_{i}\right\}$, which is denoted as:
\begin{equation}
    {c}_{spa}\left(\dot{v}^{t}_{i}\right)= \|\emph{H}_{\mathcal{G}}^{t}\left(\dot{v}^{t}_{i}\right)\cdot\theta_{spa}-\emph{p}^{spa}_{\dot{v}^{t}_{i}}\cdot\emph{H}_{\mathcal{G}}^{t}\left(\Tilde{S}^{spa}\left(\dot{v}^{t}_{i}\right)\right)\|_{2}
    ~\label{eq:spa_cons}
\end{equation}
where ${c}_{spa}\left(\dot{v}^{t}_{i}\right)$ denotes the spatial reconstruction error. $\left\|\cdot\right\|_{2}$ denotes the $l2$ norm of the vector. $\emph{H}_{\mathcal{G}}^{t}\left(\dot{v}^{t}_{i}\right)$ and $\emph{H}_{\mathcal{G}}^{t}\left(\Tilde{S}^{spa}\left(\dot{v}^{t}_{i}\right)\right)$ are node feature matrices of the master node and the spatial candidate node set respectively. $\theta_{spa}$ is a trainable projection matrix when generating the spatial hyperedge $e_{spa}\left(\dot{v}_{i}^{t}\right)\in\mathcal{E}^{t}$.
$\emph{p}^{spa}_{\dot{v}^{t}_{i}}\in\mathcal{R}^{\left(N-1\right)}$ is the trainable reconstruction coefficient vector. $S^{spa}\left(\dot{v}^{t}_{i}\right)$ is denoted as: 
\begin{equation}
S^{spa}\left(\dot{v}^{t}_{i}\right) = \left\{v|v\in\Tilde{S}^{spa}\left(\dot{v}^{t}_{i}\right),\emph{p}^{spa}_{\dot{v}^{t}_{i}}\left(v\right)>\zeta\right\} 
\label{eq:zeta}
\end{equation}

Based on the trainable vector $\emph{p}^{spa}_{\dot{v}^{t}_{i}}$ and threshold $\zeta$, we filter the candidate nodes to construct the tail node set $\mathcal{H}_T$ in a more flexible manner. We group master nodes that share the same candidate node set to form the head node set $\mathcal{H}_H$. Finally, we form directed hyperedges composed of head and tail node-sets. Similarly, temporal candidate nodes are encircled from $\emph{V}^{t-1}$ to form the temporal hyperedge. Overall, the loss of hyperedges generation in one timestamp is defined as:
\begin{equation}
\begin{split}
        \mathcal{L}_{recon}\!\! = \!\!\!\!\sum_{i=\left[1,...,N\right]}&\!\!\!\!\!\!\lambda\left(c_{spa}\!\left(\dot{v}^{t}_{i}\right)\!+\!c_{tem}\!\left(\dot{v}^{t}_{i}\right)\right)
        \!+\!\left(  \|\emph{p}^{spa}_{\dot{v}^{t}_{i}}\|_{1}+\|\emph{p}^{tem}_{\dot{v}^{t}_{i}}\|_{1}\right)\\&+{\gamma_2}\left(
    \|\emph{p}^{spa}_{\dot{v}^{t}_{i}}\|_{2}+\|\emph{p}^{tem}_{\dot{v}^{t}_{i}}\|_{2}\right)
    ~\label{eq:tem_cons}
\end{split}
\end{equation}
where $\left\|\cdot\right\|_{1}$ denotes the $l1$ norm of the vector and $c_{tem}$ denotes the reconstruction error of temporal hyperedges. $\lambda$ is the weight hyperparameter. $\gamma_2$ is the regularizing factor to balance $l1$ norm and $l2$ norm. \revises{The reconstruction error ensures accurate feature restoration, $l1$ aids feature selection, and $l2$ prevents overfitting, enabling the model to capture high-order spatio-temporal correlations in road networks.}

\subsection{Directed Hypergraph Learning}

Let $\emph{M} \in R^{N \times N}$ represents the correlation between all nodes  and the  master node $\dot{v}^{t}_{i}$ at time $t$, with entries are as below:
\begin{equation}
\emph{M}\left(v,e\left(\dot{v}^{t}_{i}\right)\right) =\left\{ 
    \begin{matrix}
      1\text{,} & v = \dot{v}^{t}_{i} \\
    \emph{p}_{\dot{v}^{t}_{i}}\!\left(\!v\!\right)\!\text{,} & v \in  \mathcal{H}_T\left(\!\dot{v}^{t}_{i}\!\right) \\
      {1}/{|\mathcal{H}_H\left(\!\dot{v}^{t}_{i}\!\right)|}  \text{,} & v \in \mathcal{H}_H\left(\!\dot{v}^{t}_{i}\!\right)  \\
      0\text{,} & \text{otherwise}
    \end{matrix}\right.
\end{equation}
\revises{where $\mathcal{H}_H$ and $\mathcal{H}_T$ denote the set of head nodes and the set of tail nodes, respectively.}

The embedding of hyperedges is aggregated by node features following: 
\begin{equation}
    \emph{E}\left(e\left(\dot{v}^{t}_{i}\right)\right) = \frac{\sum_{v\in\mathcal{V}^{t}}\emph{M}\left(v,e\left(\dot{v}^{t}_{i}\right)\right)\times\emph{H}_{\mathcal{G}}^{t}\left(v\right)}{\sum_{v\in\mathcal{V}^{t}}\emph{M}\left(v,e\left(\dot{v}^{t}_{i}\right)\right)}
    \label{eq:edge_emb}
\end{equation}

We calculate multi-head attention between the master node and the two types of hyperedge ($e_{spa}\left(\dot{v}^{t}_{i}\right),e_{tem}\left(\dot{v}^{t}_{i}\right)$). The calculation process for the weights of the two types of hyperedge is similar, and thus we obtain spatial hyperedge attention $att^{h}_{spa}\left(\dot{v}^{t}_{i}\right)$ by:
\begin{equation}
att^{h}_{spa}\left(\dot{v}^{t}_{i}\right) =  \frac{\emph{H}_{\mathcal{G}}^{t}\left(\dot{v}^{t}_{i}\right)\cdot Q\text{-}Lin^{h} \cdot\Theta_{spa}^{att}\cdot\emph{K}_{spa}^{h}\left(\dot{v}^{t}_{i}\right)^{T}}{\sqrt{d}}
\end{equation}
where
\begin{equation}
\emph{K}_{spa}^{h}\left(\dot{v}^{t}_{i}\right) = \emph{E}\left(e_{spa}\left(\dot{v}^{t}_{i}\right)\right)\cdot K\text{-}Lin^{h}_{spa}
\end{equation}

Here, for the $h$-th attention head $att^{h}\left(\dot{v}^{t}_{i}\right)$, $Q\text{-}Lin^{h}\in\mathcal{R}^{d\times\frac{d}{K}}$ is the linear transformation matrix, and $K\text{-}Lin^{h}_{spa}$ is a trainable matrix with the same dimensions. $K$ is the number of attention heads. $\Theta_{spa}^{att}\in\mathcal{R}^{\frac{d}{K}\times\frac{d}{K}}$ is trainable weight matrix. The $h$-th temporal attention $att^{h}_{tem}\left(\dot{v}^{t}_{i}\right)$ is calculated in a similar manner. The weight of the spatial hyperedge $w^{h}_{spa}\left(\dot{v}^{t}_{i}\right)$ and the temporal hyperedge $w^{h}_{tem}\left(\dot{v}^{t}_{i}\right)$ are calculated by $softmax$ normalization to $att^{h}_{spa}\left(\dot{v}^{t}_{i}\right)$ and $att^{h}_{tem}\left(\dot{v}^{t}_{i}\right)$.
\begin{equation}
\begin{split}    
\emph{Q}_{\emph{V}^{t}}\left(\dot{v}^{t}_{i}\right) = \|_{k=1}^K\Bigl(w^{h}_{spa}\left(\dot{v}^{t}_{i}\right)\times \emph{K}_{spa}^{h}\left(\dot{v}^{t}_{i}\right)\\+w^{h}_{tem}\left(\dot{v}^{t}_{i}\right)\times \emph{K}_{tem}^{h}\left(\dot{v}^{t}_{i}\right)\Bigl)
\end{split}
~\label{eq:att}
\end{equation}
where $ \|$ is the concatenation operator. We average the the node embedding of all nodes $Q_{V^t}$ to generate the graph representation of $\mathcal{G}^t$, which is represented as $Q_{\mathcal{G}^t}$. 

\subsection{Multi-agent Deep Reinforcement Learning}
\label{sssec:madrl}
PPO-Clip~\citep{PPO} modifies the surrogate objective by clipping the probability ratio, to penalize changes to the policy that
move the value of $r_t(\theta)$ away from 1. $r_t(\theta)$ is defined as:
\begin{equation}
r_t(\theta) = \frac{\pi_\theta(a_t \mid s_t)}{\pi_{\theta_{\text{old}}}(a_t \mid s_t)}
\end{equation}

The policy loss function is obtained by:
\begin{equation}
L^{clip}_{\pi}(\theta) = \hat{\mathbb{E}}_t \left[ \min \left( r_t(\theta) \hat{A}_t, \text{clip}(r_t(\theta), 1 - \epsilon, 1 + \epsilon) \hat{A}_t \right) \right]
\end{equation}
where $\epsilon$ is a hyperparameter indicating the range for clipping. $\hat{A}_t$ is an estimator of the advantage function at time $t$:
\begin{equation}
\hat{A}_t = \delta_t + (\gamma \lambda_{GAE}) \delta_{t+1} + \cdots + (\gamma \lambda_{GAE})^{T-t+1} \delta_{T-1}
\end{equation}
where the discount rate $\gamma$ is used to balance short-term and long-term rewards. $\lambda_{GAE}$ is the smoothing coefficient. The temporal difference error is calculated by:
\begin{equation}
    \delta_t = r_t + \gamma V(s_{t+1}) - V(s_t)
\end{equation} 

To realize the procedure in an end-to-end fashion, we design the loss function used in the training process as:
\begin{equation}
    \mathcal{L}_\textit{HG} = \beta\mathcal{L}_{recon}+\left(1-\beta\right)\textit{MSE}\left(\textit{MLP}\left(\emph{Q}_{\mathcal{G}^{t}}\right),\emph{y}^{t}\right)
    ~\label{eq:HG_loss}
\end{equation}
where \textit{MSE} is the mean squared error function. $\beta$ is a weight hyperparameter to balance the impact of reconstruction loss and Q-value loss. $y^t$ is the target Q-value.

\subsection{Time Complexity Analysis}
The observation embedding layer is involved with time complexity $O(Nd^2)$, where $N$ is the number of nodes, $d$ is the feature dimension. For the actor network, the time complexity is $O(Nd^2)$. In critic network, the time complexity of directed hypergraph construction is $O(Nd^2+N^2d)$. The time complexity of directed hypergraph learning is $O(Nd+NN+Nd^2+N^2d) \approx O(Nd^2+N^2d)$. The time complexity of graph readout is $O(Nd^2)$. Hence, the overall time complexity is $O(6Nd^2+2N^2d+Nd+NN) \approx O(Nd^2+N^2d)$.

\section{Experiments}
\label{sec:experiment}
\subsection{Experimental Setup}
We establish a simulation testbed based on CityFlow~\citep{CityFlow}, which is an open-source traffic simulator that supports large-scale traffic signal control. After feeding the given traffic datasets into the simulator, vehicles navigate towards their destinations according to the defined environmental parameters. This simulator assesses the road conditions based on the specified traffic signal control strategy and models the actions of each vehicle, offering the details of the traffic evolution within the road network. The main parameter configurations for our proposal are presented in Table~\ref{tab:params}.


\subsection{Dataset}

\revise{We introduce four datasets  (Table~\ref{tab:flow-details}) to evaluate the effectiveness of our proposed method: two synthetic datasets, $D_\textit{4$\times$4}$ and $D_\textit{6$\times$6}$, along with two real-world datasets, $D_\textit{Hangzhou}$ and $D_\textit{Jinan}$. The specifics of these datasets are outlined below.}

\begin{figure}[t!]
	\centering
        \begin{minipage}[b]{\columnwidth}
		\centering
		
		\includegraphics[scale=0.43]{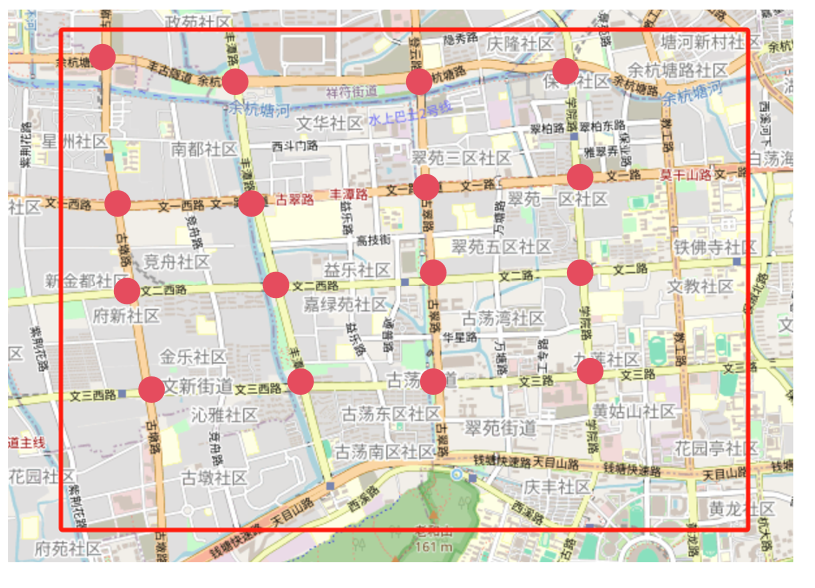}
		\subcaption{$D_\textit{Hangzhou}$}\label{}
	\end{minipage} 
	\begin{minipage}[b]{\columnwidth}
    \vspace {0.3cm}
		\centering
		\includegraphics[scale=0.4
        ]{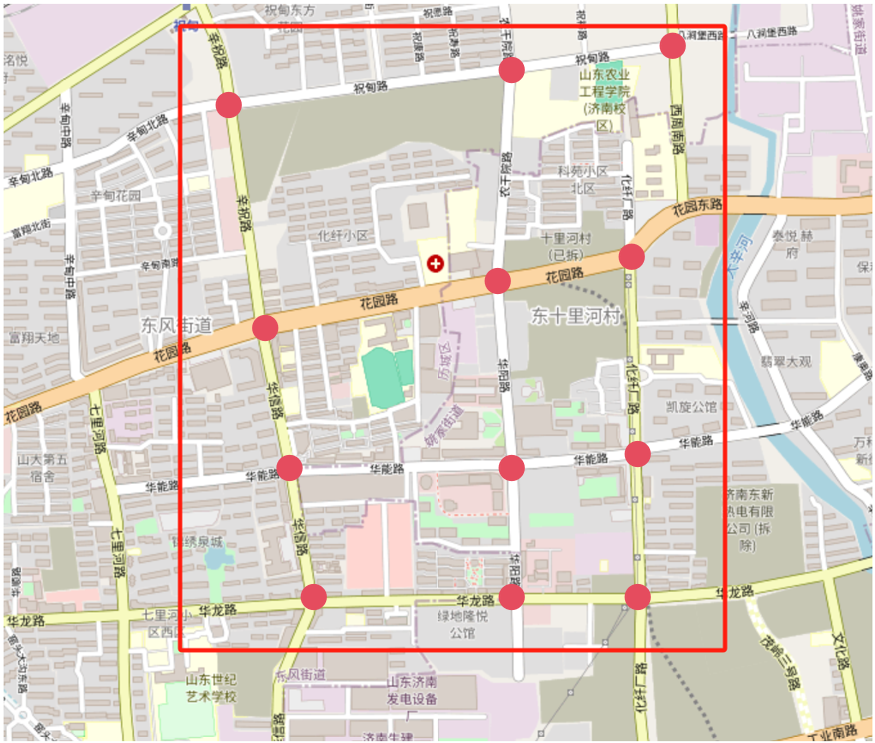}
		\subcaption{$D_\textit{Jinan}$}\label{}
	\end{minipage}
    \vspace {0.2cm}
	\caption{Road networks for real-world datasets. }	
\label{fig:Road networks}
 \vspace {1cm}
\end{figure}

 \begin{itemize}
\item 
\textbf{Synthetic Datasets}:
\revise {We examine two synthetic traffic datasets of varying scales. For the $D_\textit{4$\times$4}$ dataset, the simulator generated a total of 1,473 traffic flow data records following the Gaussian distribution. All vehicles entered and exited the network from the edge roads on the periphery. The turning ratios of the vehicles were 10\% for left turns, 60\% for going straight, and 30\% for right turns.  In the $D_\textit{6$\times$6}$ dataset, vehicles arrived uniformly, with a flow of 300 vehicles per lane per hour in the west-east  direction and 90 vehicles per lane per hour in the south-north  direction.}

\item \textbf{Real-world Datasets}:
Two real-world datasets $D_\textit{Hangzhou}$ and $D_\textit{Jinan}$~\citep{Colight} are adopted in the experiment. 
Specifically, the road layout in $D_\textit{Hangzhou}$ consists of a $4\times4$ grid located in Gudang Sub-district. In a similar manner, $D_\textit{Jinan}$ features a $3\times4$ grid in Dongfeng Sub-district of Jinan. Both datasets provide information regarding the road network configurations and the associated traffic flow details. \revise{The distribution of intersections within these two datasets is illustrated in \figurename~\ref{fig:Road networks}.}

 \end{itemize}

\begin{table}[tb]
 \caption{Main parameter settings.}
 \label{tab:params}
 \centering
 \begin{tabular}{l|l}
   \hline
   Parameter  & Value \\
   \hline 
   Batch size & 50\\
   Episodes & 100\\
   
   Learning rate & 0.0003 (actor \& critic)\\
   Replay memory ($\textit{Thres}_\textit{size}$) & 1000\\
   Optimizer & Adam\\
   Head number (K) & 1\\
   Discount rate ($\gamma$)  & 0.91\\
   Smoothing coefficient ($\lambda_{GAE}$)  & 0.86\\
   Reconstruction error ($\lambda$) & 0.001\\
   Regularizing factor ($\gamma_2$) & 0.2\\
   Clipping parameter ($\epsilon$) & 0.3 \\

   \hline
  \end{tabular}
\end{table}

\begin{table}[tb]
 \vspace {0.5cm}
 \caption{Traffic flow details of different traffic datasets.}
 \label{tab:flow-details}
 \vspace {0.1cm}
 \centering
 \begin{tabular}{l|c|c|l}
   \hline
 \multicolumn{1}{c|}{\multirow{1}{*}{Dataset}} &\multicolumn{1}{|c}{\multirow{1}{*}{Intersections}}&\multicolumn{1}{|c}{\multirow{1}{*}{Total Vehicle Flow}}
  &\multicolumn{1}{|c}{\multirow{1}{*}{Type}} \\\cline{1-4} 
  $D_\textit{Hangzhou}$ & 16 &6984& Real-world\\ 
    $D_\textit{Jinan}$ & 12 &6295 &Real-world \\
    $D_\textit{4$\times$4}$ & 16 &1473& Synthetic \\
    $D_\textit{6$\times$6}$ & 36 &3000 & Synthetic \\\cline{1-4} 

  \end{tabular}
\end{table}


\subsection{Baselines}
We use twelve methods as baselines to validate the effectiveness of our proposed \textbf{\sysname}. The baselines encompass traditional transportation methods, reinforcement learning techniques, and those that integrate graph learning. To ensure a fair comparison, we utilize the optimal parameters configurations for various datasets as specified in the original open source codes of the baseline methods. The details of the baseline methods are as below: \begin{itemize}
    \item \textbf{Fixed-time}~\citep{Koonce2008TrafficST}: Assigning a constant cycle length with a predetermined green ratio distribution across all phases.

    \item \textbf{MaxPressure}~\citep{Varaiya2013}: Controlling traffic signals by alleviating congestion on the upstream and downstream queue length with the max pressure. 

    \item \textbf{GCN}~\citep{GCN2018}: Using graph convolution neural network to extract traffic features among multiple intersections. 

    \item \textbf{PressLight}~\citep{WeiHuaKDD19}: Integrating pressure into the state and reward design of the reinforcement learning model to solve the multi-intersection signal control problems.

    \item \textbf{CoLight}~\citep{Colight}: Using a graph attention network to enhance coordination and decision-making of multi-intersection traffic signal control.
    
    \item \textbf{MPLight}~\citep{ChenAAAI2020}: A decentralized network level traffic signal control RL algorithm with parameter sharing which enables large scale application based on Presslight.

    
        \item \textbf{Attn-CommNet}~\citep{GaoICTAI21}: \revise{A reinforcement learning method that incorporates a local attention mechanism, leveraging local selection and attention between hidden layers to enhance cooperation among agents.} 
        \item{\textbf{GPLight}~\citep{GPLight}: A multi-agent reinforcement learning method that maintains a balance between accuracy and complexity in large-scale traffic signal control systems by clustering agents that share a high degree of similarity.}


  

    
    \item \textbf{SO2}~\citep{SO2}: An offline-to-online reinforcement learning method that improves Q-value estimation by perturbing the target action and boosting the frequency of Q-value updates. 


     \item \textbf{EMC}~\citep{WangTITS24}: \revise{ A multi-agent coordination method that models each intersection as an autonomous agent, using a cost function to optimize traffic signal decisions and minimize vehicle travel time through decentralized message passing.}
    \item\textbf{CrossLight}~\citep{CrossLight}: It combines offline meta-training with online adaptation to facilitate the transfer of traffic signal control strategies between different cities.
    
    
    \item \textbf{MixLight}~\citep{MixLight}: A multi-agent reinforcement learning approach that employs an executor-guide dual network to facilitate bilateral cooperation in multi-intersection TSCSs.

\end{itemize}


\begin{table}[t!]
    \centering
    \caption{The ATT performance of different methods.}
    \label{tbl:table1}
    \begin{tabular}{l|ccccccc}
        \hline
        Method [Source] & \multicolumn{4}{c}{Average travel time (secs)} \\
        
         & $D_\textit{Hangzhou}$ & $D_\textit{Jinan}$ & $D_\textit{4$\times$4}$ & $D_\textit{6$\times$6}$ \\

        \cline{1-1}
        \hline
        
        Fixed-time & 718.29 & 814.11 & 213.04 &334.68 \\
        MaxPressure & 407.17 & 343.90 & 159.56 &195.49\\
        \hline
        GCN [ITSC18'] & 709.70 & 596.67 & -& 272.14\\
        
        PressLight [KDD19']  & 416.73 & 311.39 & 232.04 &256.89 \\
        CoLight [CIKM19'] & 302.37 & 309.93 & 141.79 &173.88 \\
        MPLight [AAAI20'] &319.71 & 304.93 & 177.01 &238.27\\
        Attn-CommNet [ICTAI21'] &384.00&334.30&-&209.30\\
        GPLight [IJCAI23']  & 301.45 & 307.52 & - & -  \\
        SO2 [AAAI24'] & 333.29 & 281.02 & 135.08 &-\\
        EMC [TITS24'] &362.10&304.10&-&-\\
        CrossLight [KDD24']  & 329.98 & 279.52 & 128.73&- \\
        MixLight [TII24'] &317.90& 303.04 & - &227.99\\
        \textbf{\sysname (Ours)} & \textbf{292.56} & \textbf{277.94} &\textbf{123.51}&\textbf{166.10} \\
 
        \hline
    \end{tabular} 
    \begin{tablenotes}
\item {\scriptsize $-$} indicates results are inapplicable to certain dataset.
\end{tablenotes}
\end{table}

\begin{figure*}[t!]
\vspace {-0.3cm}
	\centering
        \begin{minipage}[b]{.485\columnwidth}
		\centering
		
		\includegraphics[width=\columnwidth]{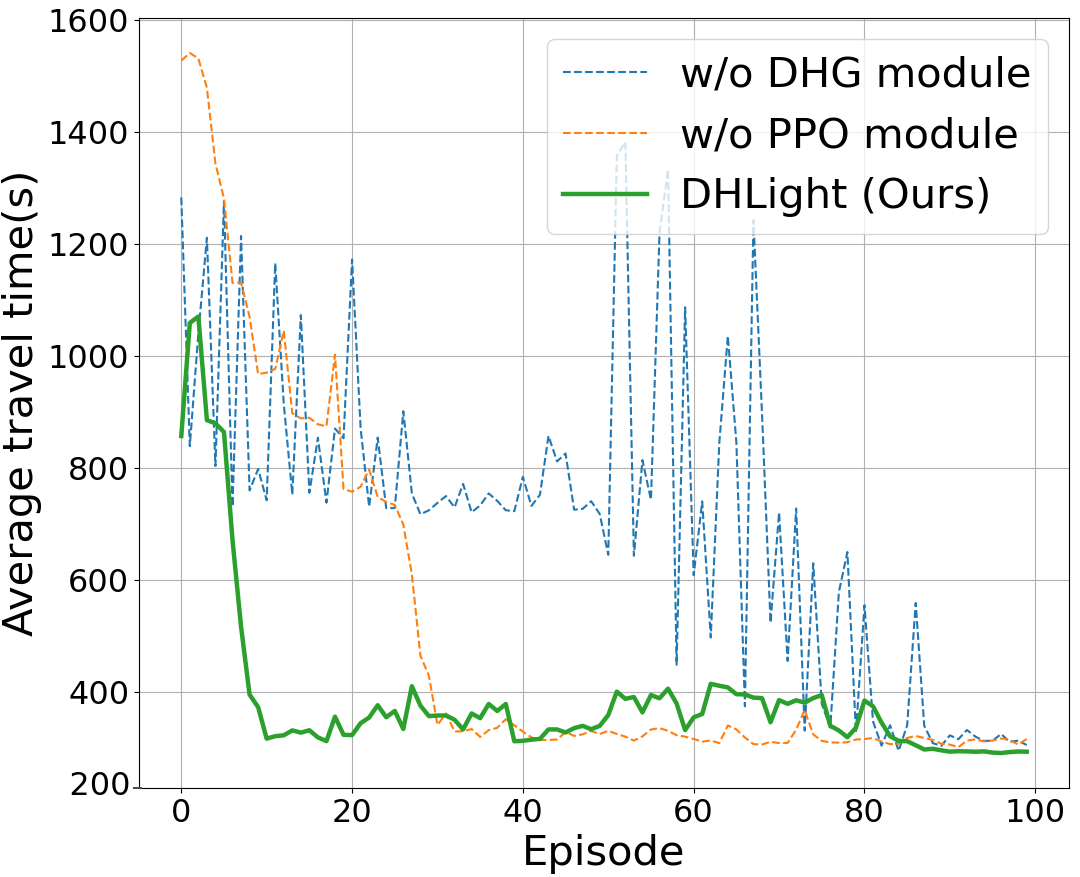}
		\subcaption{$D_\textit{Hangzhou}$}\label{fig:ablation_hangzhou}
	\end{minipage} 
	\begin{minipage}[b]{.485\columnwidth}
    \vspace {0.3cm}
		\centering
		\includegraphics[width=\columnwidth]{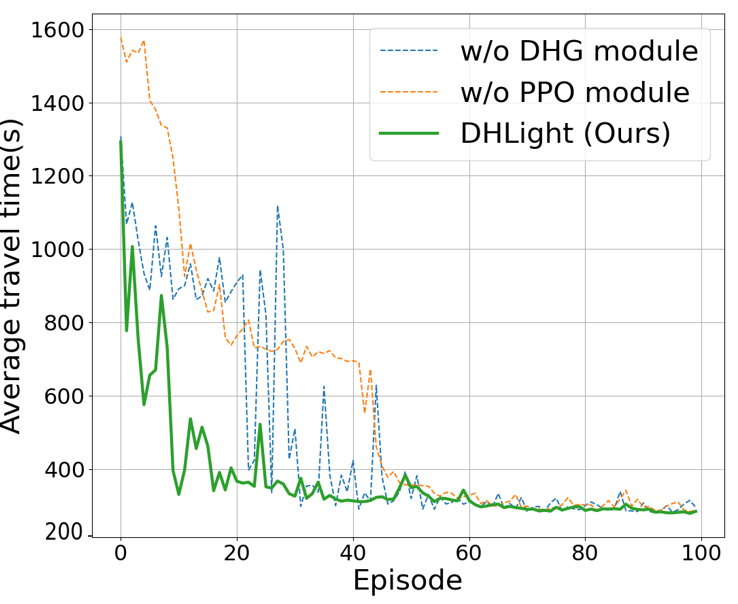}
		\subcaption{$D_\textit{Jinan}$}\label{fig:ablation_jinan}
	\end{minipage}
    	\begin{minipage}[b]{.485\columnwidth}
    \vspace {0.3cm}
		\centering
		\includegraphics[width=\columnwidth]{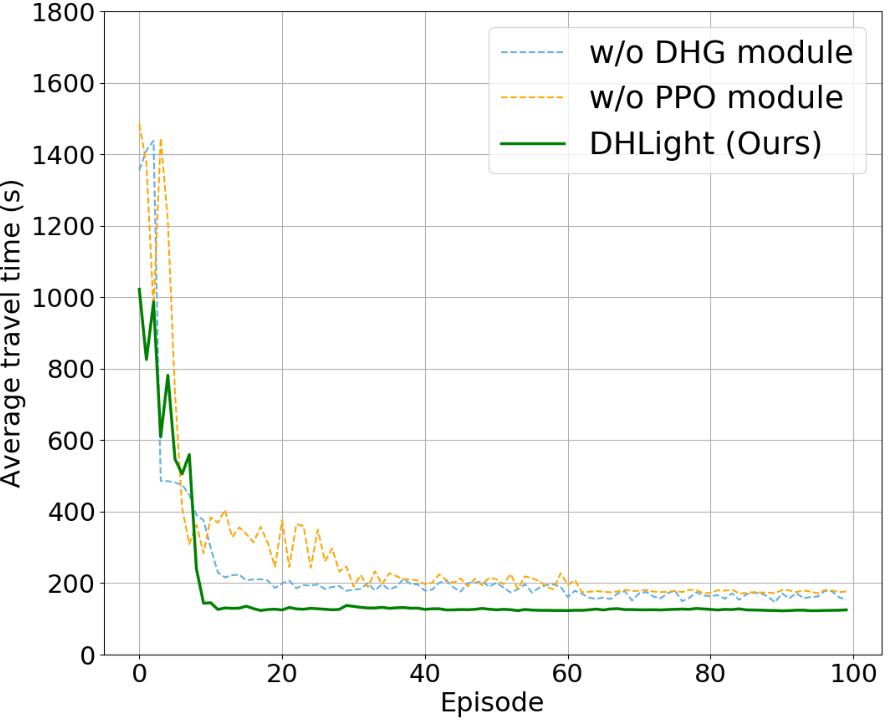}
		\subcaption{$D_\textit{4$\times$4}$}\label{fig:ablation_44}
	\end{minipage}
        	\begin{minipage}[b]{.485\columnwidth}
    \vspace {0.3cm}
		\centering
		\includegraphics[width=\columnwidth]{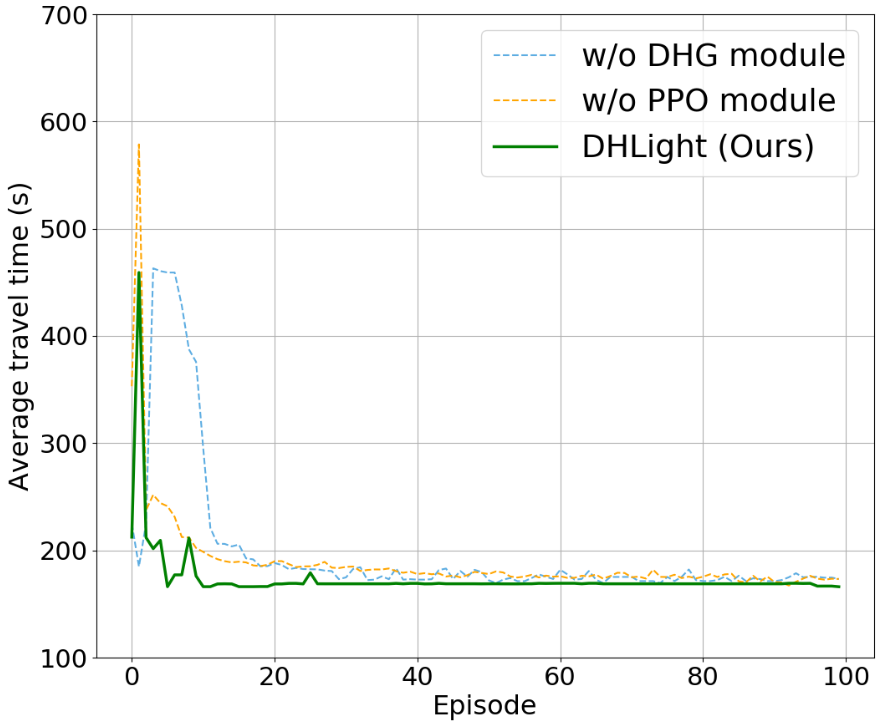}
		\subcaption{$D_\textit{6$\times$6}$}\label{fig:ablation_66}
	\end{minipage}
    \vspace {0.5cm}
	\caption{Ablation results on different datasets. }	
\label{fig:albation}
 \vspace {0.2cm}
\end{figure*}

\begin{figure*}[t]

	\centering
	\begin{minipage}[b]{.485\columnwidth}
		\centering
		\includegraphics[width=\columnwidth]{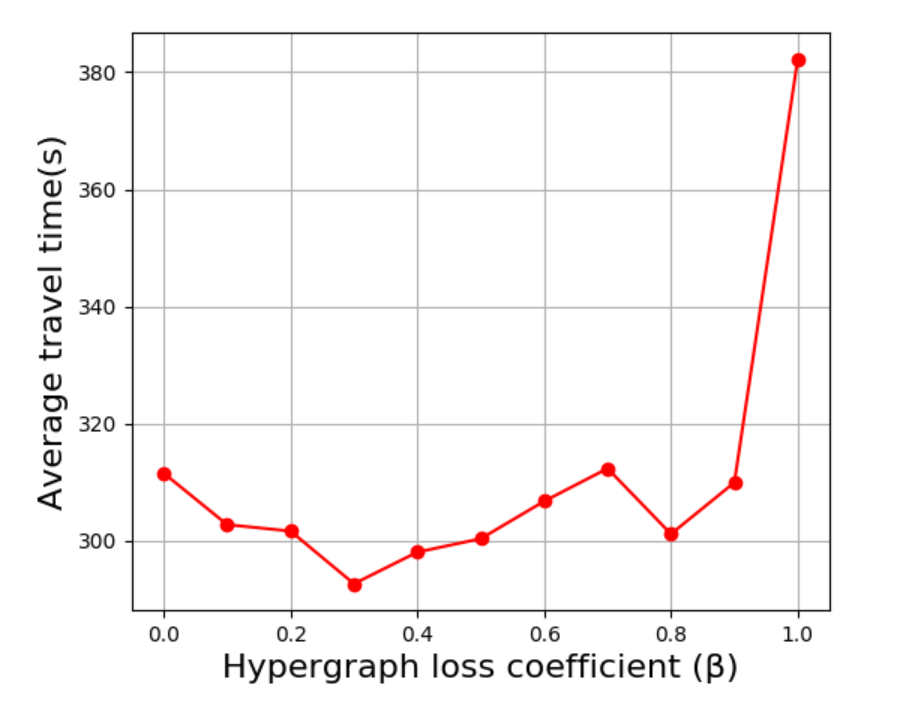}
		\subcaption{$D_\textit{Hangzhou}$, $\beta$ }\label{fig:beta_hangzhou}
	\end{minipage}
	\begin{minipage}[b]{.5\columnwidth}
		\centering
		\includegraphics[width=\columnwidth]{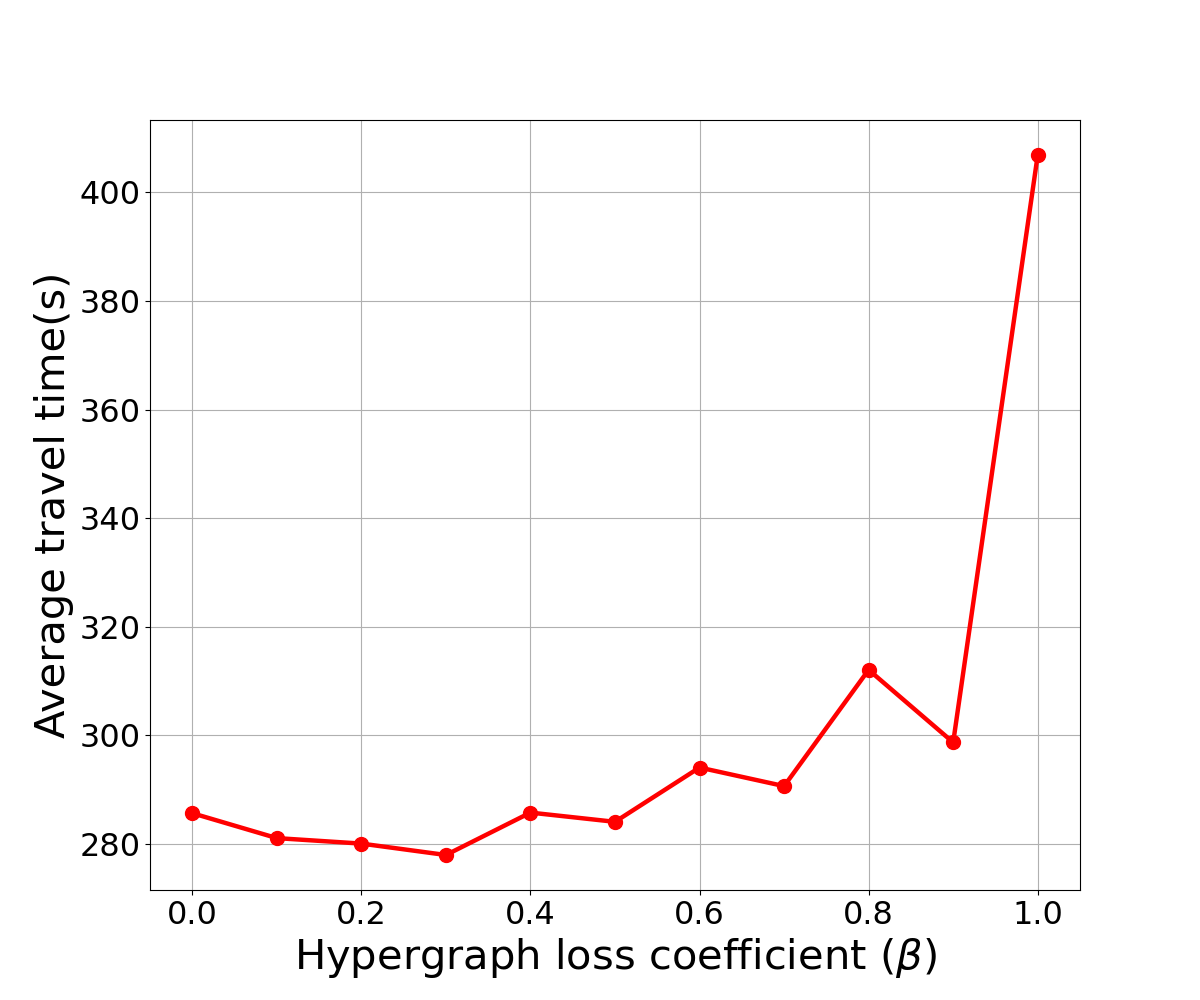}
		\subcaption{$D_\textit{Jinan}$, $\beta$}\label{fig:beta_jinan}
	\end{minipage}
	\begin{minipage}[b]{.5\columnwidth}
		\centering
		\includegraphics[width=\columnwidth]{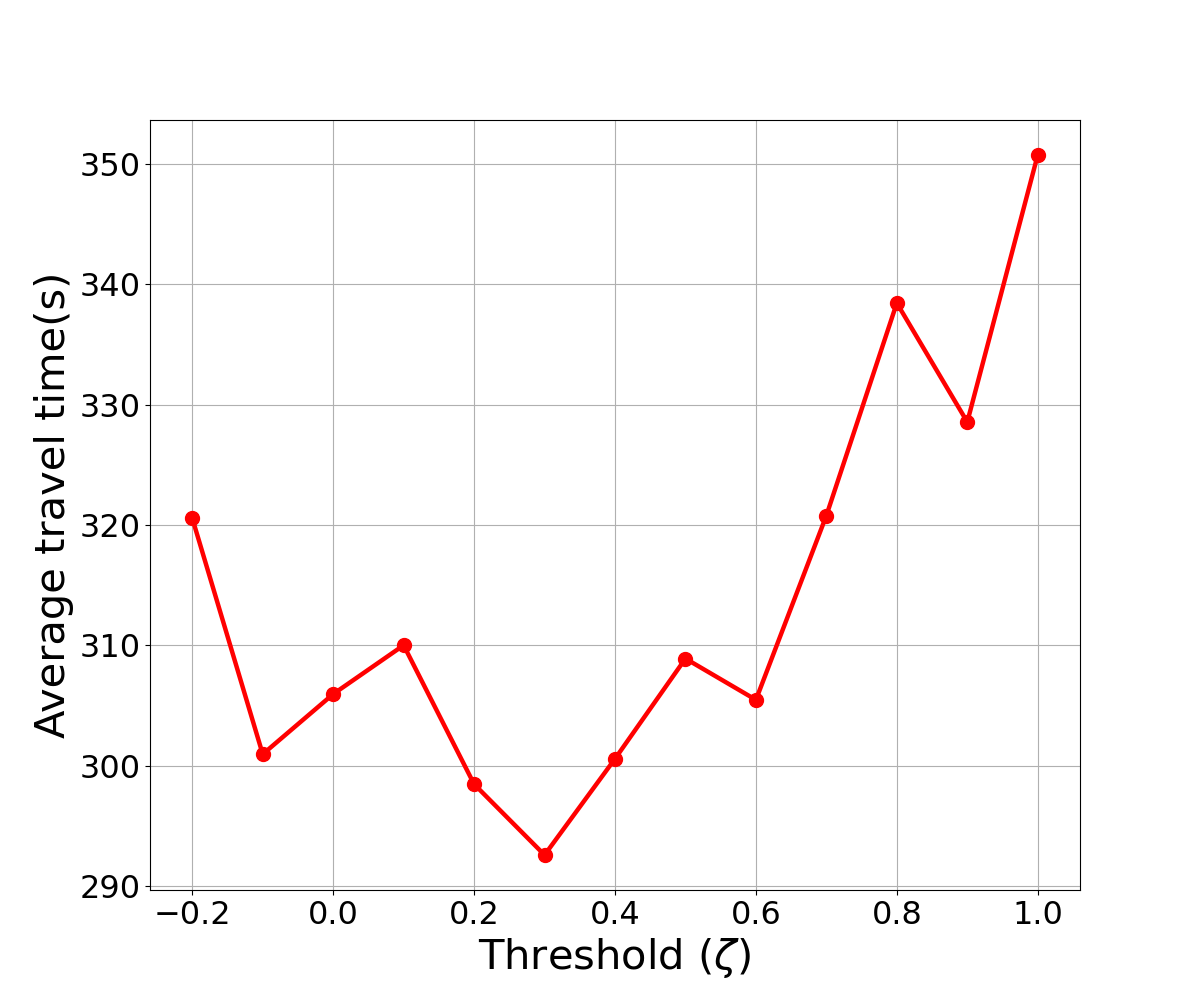}
		\subcaption{$D_\textit{Hangzhou}$, $\zeta$}\label{fig:zeta_1}
	\end{minipage}
	\begin{minipage}[b]{.5\columnwidth}
		\centering
		\includegraphics[width=\columnwidth]{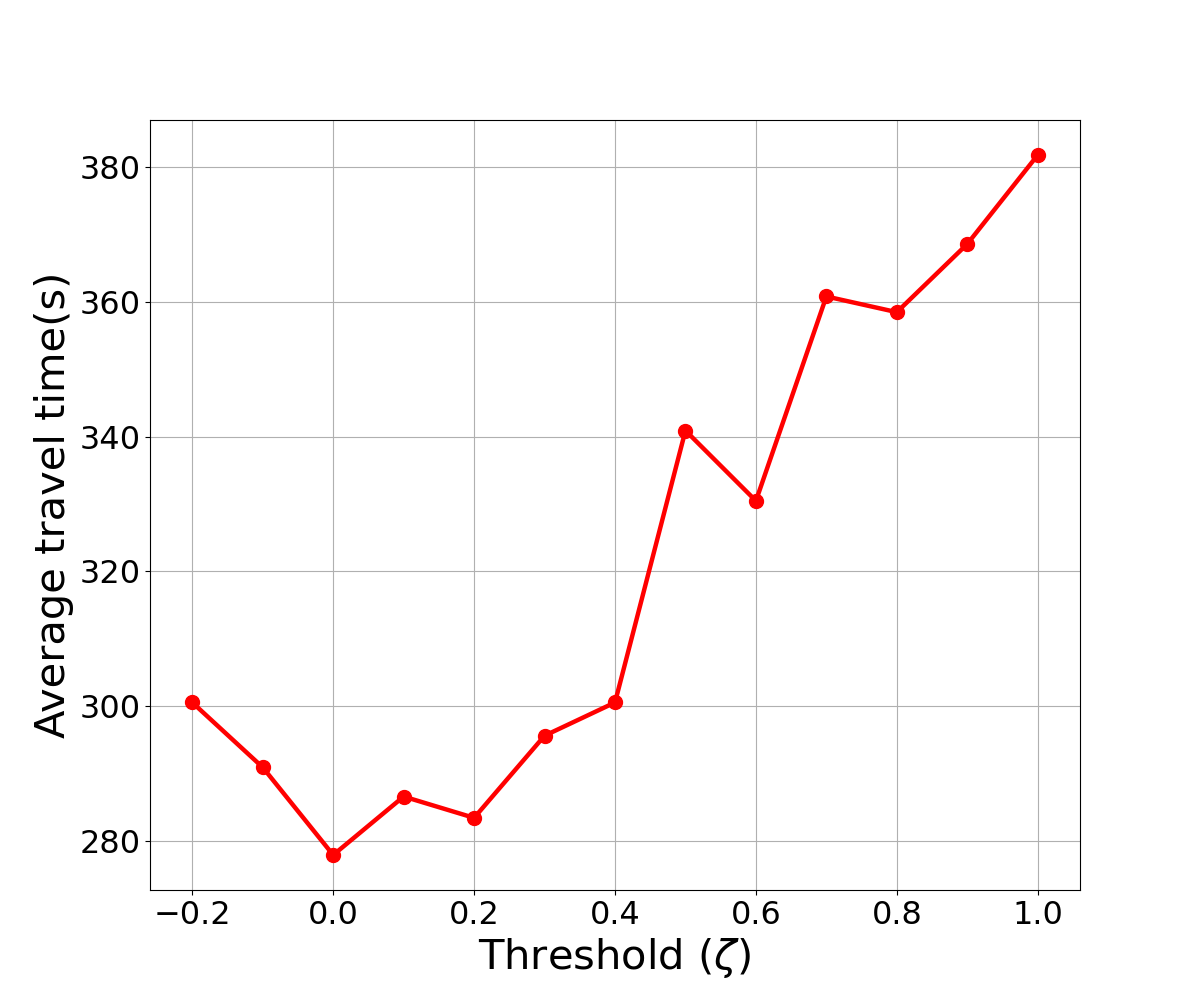}
		\subcaption{$D_\textit{Jinan}$, $\zeta$}\label{fig:zeta_2}
	\end{minipage}

     \vspace {0.5cm}
	\caption{Impact of hypergraph loss coefficient $\beta$ and threshold $\zeta$ on real-world datasets ($D_\textit{Hangzhou}$ and $D_\textit{Jinan}$).}
    \vspace{0.3cm}
	\label{fig:sensitivity}
\end{figure*}

\subsection{Performance Evaluation}

\revise{In the field of traffic signal control, Average Travel Time (ATT) is commonly used as a metric to evaluate the performance of TSCSs. ATT is defined as the average time all vehicles spend traveling between entering and exiting a designated area. This value is influenced by factors such as red and yellow traffic signals, as well as traffic congestion. A lower ATT indicates better performance of the TSCSs.}
    
Table \ref{tbl:table1} summarizes the performance of our proposed \sysname against baselines in four datasets. The results demonstrate that \sysname outperforms the baselines in minimizing ATT: 
Traditional control methods (Fixed-time and MaxPressure) rely on predetermined strategies to regulate traffic signals, which results in unsatisfactory performance in terms of ATT. In contrast, GCN and CoLight utilize GNNs to enable interactions between multiple intersections to improve overall efficiency. However, in real-world datasets, the dynamic changes in traffic flow create higher-order correlations between intersections that traditional graphs cannot fully capture. PressLight and MPLight incorporate the concept of MaxPressure at each intersection but do not account for the dynamic changes in traffic flow that introduce complex spatio-temporal attributes to the entire road network. \revise{The Attn-CommNet model incorporates a local attention mechanism to improve cooperation among agents by facilitating local selection and attention between hidden layers. However, this approach may fall short in capturing global information in complex scenarios, limiting the effectiveness of agent collaboration and ultimately affecting the model's performance.} \note{GPlight is effective in balancing accuracy and complexity through agent clustering, but it struggles to capture intricate interactions among agents in diverse traffic scenarios. }SO2 enhances Q-value estimation by perturbing the target action and increasing the frequency of Q-value updates; however, the performance of this method is constrained by its inability to fully address the inherited estimation bias and inaccurate Q-value ranking resulting from offline pretraining while applying to TSCSs. \revise{EMC models each intersection as an autonomous agent, using a cost function to optimize traffic signal decisions and minimize vehicle travel time through decentralized message passing; however, it is limited in capturing complex higher-order spatio-temporal relationships, failing to fully leverage the intricate interactions among traffic flows.} CrossLight facilitates the transfer of traffic signal control strategies between cities through offline meta-training and online adaptation; however, it inadequately models the spatio-temporal traffic differences across cities, which restricts its transferability. Similarly, while MixLight employs an executor-guide dual network to promote bilateral cooperation in multi-agent systems, it fails to sufficiently capture global spatio-temporal dependencies, potentially hindering its performance in large-scale traffic networks.

\revise{In contrast, our proposed \sysname effectively addresses the issues present in baseline methods by dynamically constructing directed hyperedges. By forming directed hyperedges that connect multiple intersections rather than being limited to actual geographical connections, \sysname  is able to capture and represent the complex directed spatio-temporal attributes inherent in TSCSs. This innovative approach enables a more comprehensive understanding of the interdependencies among various traffic nodes, allowing for deeper insights into how changes in one area can affect others. By elucidating these relationships, DHLight enhances decision-making processes and improves adaptability within dynamic traffic environments. This capability is crucial for effectively responding to real-time traffic conditions, ultimately leading to more efficient traffic management and optimized flow.}


\subsection{Ablation Study}

\figurename~\ref{fig:albation} plots the results of ablation study for our proposed \sysname. “w/o DHG module” replaces the directed hypergraph module with a traditional pairwise GAT to capture correlations between intersections.The results demonstrate that \sysname achieves satisfactory ATT performance more rapidly and smoothly compared to the traditional graph learning approach (“w/o DHG module”). \revise{This result can be attributed to the capability of directed hypergraphs to simultaneously represent relationships involving multiple intersections, which is particularly valuable in traffic networks where interactions often involve more than two intersections. By utilizing directed edges that connect multiple nodes, \sysname can model complex traffic patterns more accurately, capturing the nuances of traffic flow between various intersections. The findings  demonstrate the exceptional capability of directed hypergraphs in capturing correlations among multiple intersections. }

Similarly, “w/o PPO module” replaces the MA-PPO-based method with the ones with the Multi-Agent Soft Actor-Critic (MA-SAC). \revise{From \figurename~\ref{fig:albation}, it is clear that \sysname achieves lower and more stable ATT during the first half of the experiments, ultimately yielding comparable results by the end. The underlying reason for this phenomenon is that the MA-PPO method provides more stable policy updates during the initial training phase, leading to the development of higher-quality initial policies. This stability allows the model to quickly and effectively capture the dynamic characteristics of the traffic network, facilitating a more responsive adaptation to changing conditions. In contrast, while the MA-SAC method boasts stronger exploration capabilities, its lower quality of initial policy adversely impacts ATT performance in the early stages of training. The exploratory nature of MA-SAC can lead to suboptimal decision-making as it navigates uncharted areas of the policy space, hindering early performance compared to the more stable and effective approach of MA-PPO. Overall, the advantages of the MA-PPO algorithm lies in its ability to formulate an effective initial policy, which is crucial for efficiently addressing complex traffic dynamics.}

\subsection{Sensitivity Analysis}

\figurename~\ref{fig:sensitivity} investigates the impact of the hyperparameter $\beta$ and $\zeta$ in Equation~\ref{eq:HG_loss} and Equation~\ref{eq:zeta}, respectively. $\beta$ is the weight coefficient assigned to the hyperedge construction losses within the model. This coefficient plays a pivotal role in balancing the contributions of different loss components during the training process, thereby influencing the overall effectiveness of hyperedge formation. \revise{The findings reveal that the optimal value of $\beta$ for both $D_\textit{Hangzhou}$ and $D_\textit{Jinan}$ is identified as 0.3.} As $\beta$ increases, the proportion of loss associated with the dynamic construction of directed hypergraphs also escalates. This indicates that $\beta$ is crucial in balancing the trade-off between the complexity of hypergraph construction and the overall performance of the model. 

Meanwhile, the threshold $\zeta$ serves as a key parameter in the dynamic construction of the hypergraph, determining which candidate nodes will be included in the hyperedges. 
 The candidate nodes whose correlation with the master node exceeds the threshold $\zeta$ will be incorporated into the hyperedges. An increase in threshold $\zeta$ leads to a decrease in the number of candidate nodes incorporated into the hyperedges. The results show that the optimal value of $\zeta$ is 0.3 for $D_\textit{Hangzhou}$ and 0.0 for $D_\textit{Jinan}$.  This reduction adversely affects the ability to effectively capture the spatiotemporal correlations between intersections, ultimately degrading the performance of ATT. \revise{Similar trends are recorded when testing on two synthetic datasets  $D_\textit{4$\times$4}$ and $D_\textit{6$\times$6}$ (We omit the details due to space limitation).}

\section{Conclusion}\label{sec:conclusion}
In this paper, we propose a multi-agent policy-based framework \sysname that incorporates a directed hypergraph module to capture the correlation in road traffic by dynamically constructing spatio-temporal hyperedges. This innovative approach allows for a deep comprehension of the directed spatio-temporal relationships between multiple intersections and their traffic signals, enabling the obtained model to adapt to varying traffic conditions in real time. Extensive experimental results demonstrate the superiority of \sysname against the baselines in traffic signal control. Our future work is to enhance the robustness of our method by studying the data cleaning schemes to mitigate  the impact of missing or noisy existing in the collected data.

\begin{ack}
This work was supported in part by the National Natural Science Foundation of China (Grant No. 62472332) and the Natural Science Foundation of Shanghai (No.25ZR1402511).
\end{ack}


\bibliography{mybibfile}

\end{document}